\documentclass[journal]{IEEEtran}
\pdfoutput=1
\usepackage{adapt}
\usepackage{eso-pic}

\IEEEoverridecommandlockouts 
\begin{document}
        \title{Adaptive Basis Function Selection for Computationally Efficient Predictions}
	\author{
		Anton Kullberg, Frida Viset, Isaac Skog,~\IEEEmembership{Senior Member,~IEEE}, %
        and Gustaf Hendeby,~\IEEEmembership{Senior Member,~IEEE}.\\%
	\thanks{\noindent This work was partially supported by the Wallenberg AI,
		Autonomous Systems and Software Program (\textsc{WASP}) funded
		by the Knut and Alice Wallenberg Foundation.}
        \thanks{Anton Kullberg and Gustaf Hendeby are with Link\"oping University, Sweden. (email: firstname.lastname@liu.se). Frida Viset is with Delft University of Technology, Netherlands. (email: F.M.Viset@tudelft.nl). Isaac Skog is with KTH Royal Institute of Technology, Sweden. (email: skog@kth.se).}
        \thanks{Source code available: \url{https://github.com/AOKullberg/adaptive-bf-selection}}
  }
\maketitle

\begin{abstract}
\gls{bf} expansions are a cornerstone of any engineer's toolbox for computational function approximation which shares connections with both neural networks and Gaussian processes.
Even though \gls{bf} expansions are an intuitive and straightforward model to use, they suffer from quadratic computational complexity in the number of \glspl{bf} if the predictive variance is to be computed.
We develop a method to automatically select the most important \glspl{bf} for prediction in a sub-domain of the model domain.
This significantly reduces the computational complexity of computing predictions while maintaining predictive accuracy.
The proposed method is demonstrated using two numerical examples, where reductions up to 50--75\% are possible without significantly reducing the predictive accuracy.
\end{abstract}


\IEEEpubid{}
\glsresetall
\AddToShipoutPictureBG*{%
	\put(0,20){
		\hspace*{\dimexpr0.075\paperwidth\relax}
		\parbox{.84\paperwidth}{\footnotesize ~\copyright2024 IEEE. Personal use of this material is permitted. Permission from IEEE must be obtained for all other uses, in any current or future media, including reprinting/republishing this material for advertising or promotional purposes, creating new collective works, for resale or redistribution to servers or lists, or reuse of any copyrighted component of this work in other works.}%
}}

\section{Introduction}\label{sec:introduction}
\IEEEPARstart{T}{he} focus of this paper is on \gls{bf} expansion models, as these are prevalent in a wide variety of disciplines~\cite{pillonetto_distributed_2018, chen_long-term_2019, torroba_online_2023, berntorpBayesianTireFrictionLearning2019, kokScalableMagneticField2018,vetterliWavelets2001,jangChannelSeparation2003,wangGalactic2020,petersenNbodyintegration2022}.
The canonical \gls{bf} model is given by 
\begin{equation}\label{eq:bfmodel}
    f(x) = \sum_{i=1}^L \phi_i(x) \theta_i,
\end{equation}
where $\phi_i$ is \gls{bf} $i$ and $\theta_i$ is its corresponding weight.
Note that $\phi_1,\dots,\phi_L$ may be interpreted as the bases for $f(\cdot)$.
For many applications, such as magnetic field mapping, weather forecasting, etc., the domain of the function $f(\cdot)$ can be vast, such that the number of \glspl{bf} $L$ needs to be very large to represent the underlying phenomena \cite{visetSpatiallyScalableRecursive2023,kullbergOnlineJointState2021,heatonCaseStudyCompetition2018}.
The downside of this is that the predictive model can become very computationally demanding to use. 
For instance, magnetic fields modeled with \gls{bf} expansions can be used to improve indoor navigation~\cite{kokScalableMagneticField2018}, but unless the domain is split into patches to remedy the scaling issues, the online computational requirements scale quadratically with $L$. 
Equivalent models can also be used for multi-agent navigation and motion planning in large nonlinear fields, but the required communication between each pair of agents at each time instance also scales quadratically with $L$~\cite{jang_multi-robot_2020, viset_distributed_2023}.
In other applications, such as tire friction modeling in autonomous vehicles \cite{berntorpBayesianTireFrictionLearning2019}, $L$ must usually be kept very small so that the computational budget is not exceeded at the expense of the predictive accuracy.

The aforementioned cases are limited for computational reasons, \ie, computing predictions becomes computationally expensive as $L$ grows larger.
One solution to remedy this, typically done a priori, is to reduce $L$.
This results in a reduction of the model's representative power, effectively restricting the function space that it spans.
Another solution is to tailor the selection of the \glspl{bf} $\phi_i$ to the particular problem at hand, such that $L$ can be kept small.
This requires specialized domain knowledge, which is not always trivial nor feasible to obtain.
A third alternative is to consider so-called compact \glspl{rbf} \cite{buhmannRadialBasisFunctions2003,buhmannRadialFunctionsCompact1998,visetSpatiallyScalableRecursive2023,kullbergOnlineJointState2021} that exhibit favorable computational properties at the cost of representational power.

The aforementioned alternatives deal with this computational problem as an \emph{a priori} problem.
In contrast, we will approach the model reduction problem as an \emph{a posteriori} problem.
To that end, we seek to determine which \glspl{bf} are necessary to represent a given \gls{bf} expansion up to some tolerance, effectively reducing the model order purely for predictive purposes.
Hence, the weights of the \gls{bf} expansion are already learned and our task is to compress the model at prediction time so that it becomes computationally cheaper to use, without sacrificing too much accuracy.
That is, the model order $L$ used during the learning phase is still kept large, but the \emph{predictive} model uses only a few $\phi_i$ corresponding to the most important \glspl{bf} for the specific test input at hand.
We remark that while hyperparameter optimization, in, \eg, \gls{rbf} expansion models \cite{zhongRBFLocationOptimization2016,cammarasanaRBFoptimization2021}, may be interpreted as a type of \gls{bf} selection, this is done during training.
It thus requires training data and can furthermore not adapt to the test input.

In this paper, we present an approach for prediction time \gls{bf} selection that
\begin{enumerate}
    \item Avoids using specialized domain knowledge to select \glspl{bf} a priori and defers that selection to a posteriori.
    \item Only uses the \glspl{bf} that are important for the test input, thereby adapting the predictive model to the test input while reducing the computational complexity as compared to using the full model.
    \item Does not require additional data for selecting relevant \glspl{bf}, in contrast to previous approaches.
\end{enumerate}
We require a \gls{bf} selection process that has low computational complexity, such that the overall complexity of computing the predictions is lower than that of the full model.
Test input adaptivity is handled by optimally reducing the model in a sub-domain $\Omega$, that covers the test input of interest at prediction time.
The approach is developed for a general \gls{bf} expansion model and extended to a commonly used sparse \gls{gp} approximation.
Thus, our contribution is a computationally efficient, data-free, method of prediction time \gls{bf} selection that adapts to the test input.

\section{Basis Function Selection}\label{sec:selection}
Without loss of generality, we assume that $\phi_1,\dots,\phi_L$ are ordered in some way, such that reducing $L$ corresponds to removing particular components of the \gls{bf} expansion \cref{eq:bfmodel}.
Conceptually, this may be thought of as a Fourier series expansion where the removal of $\phi_i$ corresponds to removing some particular frequency in the representation.
We first describe our selection procedure in the deterministic case and then move to the stochastic case where $\theta\sim p(\theta)$.

\subsection{Adaptive basis function selection}\label{subsec:adaptiveselection}
The problem is to select a subset of the bases of the model $f(x)$ and use this subset as our predictive model, \ie, 
\begin{equation}\label{eq:approxbfexpansion}
    \hat{f}(x) = \sum_{j\in \mathcal{J}} \phi_j(x)\theta_j.
\end{equation}
Here, $\mathcal{J}$ is a set of indices, corresponding to the particular $\phi_j$, that best represents the original model in some sense.
Note that $\lvert \mathcal{J} \rvert = n_J < L$.

\subsubsection{Finding $\mathcal{J}$}
Formally, in the deterministic sense, finding $\hat{f}(x)$ can be formulated as an optimization problem given by
\begin{subequations}\label{eq:optproblem}
\begin{alignat}{2}
    &\underset{\mathcal{J}}{\min} &\qquad& \lVert f(x) -\hat{f}(x) \rVert^2_2 \triangleq \mathcal{L}\label{eq:norm}\\
    &\text{s.t.} && \lvert \mathcal{J} \rvert = n_J < L.
\end{alignat}    
\end{subequations}
Here, $f(\cdot)$ and $\hat{f}(\cdot)$ are given by \cref{eq:bfmodel} and \cref{eq:approxbfexpansion}, respectively.
Further, 
$
\lVert g(x) \rVert^2_2=\int_\Omega \lvert g(x) \rvert^2 dx.
$
Note that \cref{eq:optproblem} does \emph{not} imply that the estimator \cref{eq:approxbfexpansion} is going to be \gls{rmse} optimal with respect to the true underlying function.
It only means that $\hat{f}(\cdot)$ will be a good estimator of $f(\cdot)$.
Hence, there may be another model in the model class of $\hat{f}(\cdot)$ that performs better than $\hat{f}(\cdot)$, but finding this would require access to the training data.

The problem \cref{eq:optproblem} is reminiscent of a subset/feature selection problem \cite{johnSubsetSelection1994,zhuPolynomialAlgorithmBestsubset2020,venkateshFeatureSelection2019}.
Typical subset selection problems are solved by either greedy inclusion based on some heuristic or essentially by regularizing the optimization problem through, \eg, the \gls{lasso} \cite{tibshiraniLASSO1996}.
Further, typical solutions are focused on finding the best possible predictive model given a particular cardinality of the set $\mathcal{J}$.
Thus, the general problem is 
\begin{enumerate*}[label=\roman*)]
    \item training data dependent
    \item requires a fixed cardinality
    \item not test-input dependent
\end{enumerate*}.
There are subtle differences between the general subset selection problem and \cref{eq:optproblem} that allow us to 
\begin{enumerate*}[label=\roman*)]
    \item solve \cref{eq:optproblem} efficiently without access to the training data
    \item perform automatic cardinality selection
    \item adapt the predictive model to the test data
\end{enumerate*}.


To find the solution to \cref{eq:optproblem}, notice that $\mathcal{L}$ can be written as
\begin{multline}
    \mathcal{L} = 
    \bigg\lVert \sum_{i=1}^L\phi_i(x)\theta_i - \sum_{j\in\mathcal{J}}\phi_j(x)\theta_j \bigg\rVert^2_2 \\
    = \bigg\lVert \sum_{j\notin \mathcal{J}} \phi_j(x)\theta_j \bigg\rVert^2_2 
    = \int_{\Omega} \big\lvert \sum_{j\notin \mathcal{J}} \phi_j(x)\theta_j \big\rvert^2 ~dx.
\end{multline}
The second step follows by definition and corresponds to the norm of the omitted components.
The third step is simply the definition of the norm over some domain $\Omega$.
Note that the domain $\Omega$ is a user-defined choice, essentially telling us where we want the model to be well-approximated.
Now, we can bound the integrand by
\begin{equation}
    \Big\lvert \sum_{j\notin \mathcal{J}} \phi_j(x)\theta_j \Big\rvert^2 
    \leq 
    \sum_{j\notin \mathcal{J}} \big\lvert \phi_j(x)\theta_j \big\rvert^2 
    =
    \sum_{j\notin \mathcal{J}} \big\lvert \phi_j(x) \big\rvert^2 \big\lvert \theta_j \big\rvert^2,
\end{equation}
by using the triangle inequality.
Thus, we can bound $\mathcal{L}$ by
\begin{multline}\label{eq:Lbound}
    \mathcal{L} 
    \leq \sum_{j\notin \mathcal{J}} \int_{\Omega} 
    \lvert \phi_j(x)\rvert^2\lvert\theta_j\rvert^2~dx
    = \sum_{j\notin \mathcal{J}} \int_{\Omega} \lvert \phi_j(x)\rvert^2~dx \lvert\theta_j\rvert^2.
\end{multline}
We can use this bound to select important \glspl{bf} on a particular domain $\Omega$, which may be a subset of the domain of the \gls{bf} expansion.
This allows us to adaptively select the \glspl{bf} that contribute the most to the prediction \emph{at} the particular test input we are interested in.

For models where $\int_{\Omega} \lvert \phi_j(x)\rvert^2~dx \approx \int_{\Omega} \lvert \phi_i(x)\rvert^2~dx, \forall i, j$ the bound can be further simplified to
\begin{equation}\label{eq:simplifiedbound}
    \mathcal{L} \leq C\sum_{j\notin \mathcal{J}} \lvert\theta_j\rvert^2,
\end{equation}
where $C=\int_\Omega \lvert \phi_j(x)\rvert^2~dx$.
Thus, in this case, to minimize $\mathcal{L}$, we can simply select the \glspl{bf} with the largest absolute weights.

The bounds \cref{eq:Lbound,eq:simplifiedbound} can be used in two ways.
Firstly, given a cardinality $n_J$, they allow us to quickly choose the $n_J$ \glspl{bf} contributing most to the accuracy of the predictive model at the test input at hand.
Secondly, they allow us to adaptively select the number of \glspl{bf} used when predicting in a domain $\Omega$ such that the approximation error is smaller than \cref{eq:Lbound}.
To be more precise, assume w.l.o.g. that $\theta$ is sorted such that $\lvert\theta_1\rvert < \lvert\theta_2\rvert < \dots < \lvert\theta_L\rvert$.
Starting from $i=1$, we can then sum $\lvert\theta_i\rvert$ until \cref{eq:Lbound} reaches some predefined threshold.
At that point, we simply choose the remaining $\theta_i$ as the reduced basis.
We remark that \cref{eq:Lbound,eq:simplifiedbound} can be used for many different \gls{bf} expansion models.
The requirement for \cref{eq:Lbound} is simply that we can evaluate $\int_\Omega\lvert\phi_j(x)\rvert^2~dx$, whereas \cref{eq:simplifiedbound} applies generally as long as $\int_{\Omega} \lvert \phi_j(x)\rvert^2~dx \approx \int_{\Omega} \lvert \phi_i(x)\rvert^2~dx, \forall i, j$.

\subsubsection{Probabilistic solution}
In the probabilistic case, where $\theta\sim\mathcal{N}(m, S)$, we substitute $\mathcal{L}$ for $\mathbb{E}_p[\mathcal{L}]$.
The delta method can then be used to approximate the expectation as
$\mathbb{E}_p[\mathcal{L}(\theta)]\approx \mathcal{L}(\mathbb{E}_p[\theta])=\mathcal{L}(m)$.
Thus, the problem turns into selecting the \glspl{bf} with the largest absolute mean.
Obviously, this does not consider the quality of the approximate predictive variance.
We nevertheless pursue this strategy as an approximate solution to the corresponding probabilistic \gls{bf} selection problem.
Empirically, as shown in \cref{sec:examples}, this still works well.
In this probabilistic setting, the \gls{bf} selection process is even more beneficial than in the deterministic case, as computing the predictive variance is usually an $\mathcal{O}(L^2)$ operation.

\subsubsection{Discussion}
The considered problem is closely related to offline knowledge distillation, a type of model compression, that typically attempts to find a more compact representation of a large neural network model \cite{gouKnowledgeDistillation2021,buciluaModelCompression2006}.
However, know\-ledge distillation typically requires access to a dataset such that the reduced model can be trained to mimic the original model on particular data.
In contrast, we avoid using data at the cost of a suboptimally reduced model, as we essentially throw away information from \glspl{bf} we do not select.
As such, the method can be seen as a type of pruning, also common in neural network model compression \cite{liModelCompression2023,hossainComputationalComplexity2023}.
Pruning was pioneered in \cite{lecunOptimalBrainDamage1989}, where weights of a neural network were removed using Hessian information, again requiring access to a training dataset.
Here, the information loss can theoretically be remedied by projecting the whole function $f(\cdot)$ onto the reduced bases.
However, as this is computationally intensive ($\mathcal{O}(L^3)$), it is not pursued further here.

\subsection{Hilbert--space GP dual selection}
Here, we consider a special case of \cref{subsec:adaptiveselection} applied to a commonly used sparse \gls{gp} approximation.

\subsubsection{\gls{gp} approximation}
A standard \gls{gp} is a collection of random variables, any finite number of which have a joint Gaussian distribution \cite{rasmussen2005gaussian}.
Formally, we denote a \gls{gp} by $f \sim \mathcal{GP}(0, \kappa(x, x'))$, where $\kappa(x,x')$ is the kernel, representing the covariance between the pair of inputs $x$ and $x'$.
The standard conjugate \gls{gp} assumes independent observations of the latent process values, \ie, $y \sim \mathcal{N}(f, \sigma^2I)$.
The predictive distribution of the \gls{gp} is then given by $\mathcal{N}(f^*;~\mu^*,V^*)$, where%
\begin{subequations}
    \begin{align}
        \mu^*\triangleq&~ \mathbb{E}[f^*|\vec{y}] = K_{*f}(K_{ff} + \sigma^2I)^{-1}\vec{y}\\
        V^*\triangleq&~ \mathrm{var}[f^*|\vec{y}] = K_{*f}(K_{ff} + \sigma^2I)^{-1}K_{f*},
    \end{align}
\end{subequations}
where $[K_{ab}]_{ij}=\kappa([X^a]_i, [X^b]_j)$ and $X^*$ and $X^f$ are the collection of test and training inputs, respectively.

Due to the inversion of $(K_{ff}+\sigma^2I)$, a standard \gls{gp} has a computational complexity of $\mathcal{O}(N^3)$, where $N$ is the number of data points.
Approaches to remedy this have been studied extensively, see, \eg, \cite{csatoGaussianProcessesIterative2002,quinonero-candelaUnifyingViewSparse2005,titsiasVariationalLearningInducing,wilsonKernelInterpolationScalable,hensmanGaussianProcessesBig2013,izmailovScalableGaussianProcesses2018, solinHilbertSpaceMethods2020}.
Here, we consider one of these approaches, namely the \gls{hgp} \cite{solinHilbertSpaceMethods2020}, as it has seen widespread use \cite{berntorpOnlineBayesianInference2021,kokScalableMagneticField2018,svenssonComputationallyEfficientBayesian2016a}.
In the \gls{hgp}, the kernel matrix $K_{ff}$ is approximated by an approximate eigenvalue decomposition, \ie, $K_{ff} \approx \Phi \Lambda \Phi^\top$, where $\Phi$ correspond to the eigenvectors and $\Lambda$ the corresponding eigenvalues; see \cite{solinHilbertSpaceMethods2020} for details.

The predictive distribution of the \gls{hgp} is then given by \cite{solinHilbertSpaceMethods2020}
$$
\mu^* = \Phi_*^\top \Sigma \Phi^\top \vec{y},\quad
    V^* = \sigma^2 \Phi_*^\top \Sigma \Phi_*,
$$
where $\Sigma = (\Phi^\top\Phi + \sigma^2\Lambda^{-1})^{-1}$.
This can be recognized as a \gls{bf} expansion model with posterior over the weights given by
$$
p(\theta|\vec{y}) = \mathcal{N}(\Sigma\Phi^\top\vec{y}, \sigma^2\Sigma)\triangleq \mathcal{N}(m, S).
$$
Hence, again, we can find a reduced approximate model by choosing \glspl{bf} according to their absolute mean through \cref{eq:simplifiedbound}.

\subsubsection{Dual parametrization}
It is sometimes beneficial to parametrize the \gls{hgp} using the \q{dual} parametrization
$$
\alpha = \Phi^\top \vec{y},\quad
B = \Phi^\top \Phi,
$$
as these are the only factors in the posterior predictive distribution that depend on the data.
Further, both $\alpha$ and $B$ have an additive structure, \eg, 
\begin{equation}\label{eq:secondorderdual}
[B]_{ij} = \sum_{n=1}^N\phi_i(x_n)\phi_j(x_n),    
\end{equation}
meaning that the inclusion of a new data point is extremely simple.
However, the dual parametrization means that the \gls{bf} selection process is not as simple, since we would first need to convert the posterior from the $(\alpha,B)$ parametrization to the $(m,S)$, which is an $\mathcal{O}(L^3)$ operation, defeating the whole purpose to begin with.
To see this, consider \cref{eq:Lbound} and write
\begin{multline}
    \mathbb{E}\bigg[ \sum_{j\notin\mathcal{J}} \lvert\theta_j\rvert^2 \bigg] 
    \approx \sum_{j\notin\mathcal{J}}\lvert m_j \rvert^2 
    = \\ \sum_{j\notin\mathcal{J}}\Big\lvert \sum_{i=1}^n [\Sigma]_{ji}\alpha_i \Big\rvert^2 
    \leq \sum_{j\notin\mathcal{J}}\sum_{i=1}^n \big\lvert [\Sigma]_{ji}\alpha_i \big\rvert^2.
\end{multline}
This obviously requires $\Sigma$, which costs $\mathcal{O}(L^3)$ to compute.
One may also consider using, \eg, a truncated singular value decomposition of $B$ to select important components, but this also bears with it a computational complexity of $\mathcal{O}(L^3)$.
Instead, we propose to resolve this by only considering the diagonal elements of $B$ such that
$$
[\Sigma]_{ii}\approx \frac{1}{[B]_{ii} + \sigma^2/\lambda_i},\quad [\Sigma]_{ij}\approx 0,~ i\neq j
$$
such that computing $\Sigma$ becomes an $\mathcal{O}(L)$ operation.
Under the interpretation of $B$ as the precision matrix of a \gls{bf} expansion, this corresponds to neglecting any mutual information between \glspl{bf}.
With this approximation, the bound becomes
\begin{equation}\label{eq:dualbound}
\sum_{j\notin\mathcal{J}}\sum_{i=1}^n \big\lvert [\Sigma]_{ji}\alpha_i \big\rvert^2
= \sum_{j\notin\mathcal{J}} \big\lvert [\Sigma]_{jj}\alpha_j \big\rvert^2.   
\end{equation}
Note that we only use the diagonal of $B$ for \gls{bf} selection, \emph{not} for the final predictive distribution, where we use the full $B$.
This results in a suboptimal selection strategy but we still expect it to perform well.
This is motivated by the fact that, in the \gls{hgp}, the \glspl{bf} are orthonormal and the inner product between $\phi_i$ and $\phi_j$ is given by
$$
\langle \phi_i, \phi_j \rangle = \int_\Omega \phi_i(x)\phi_j(x) dx = \delta_{ij}.
$$
Hence, $[B]_{ij}$, see \cref{eq:secondorderdual}, can be viewed as an unweighted stochastic estimate of $\langle \phi_i, \phi_j \rangle$ based on the data.
As long as the data distribution sufficiently covers the domain of the \glspl{bf}, then $[B]_{ij} \to 0$, as $N\to\infty$, if $i\neq j$.

\section{Numerical Examples}\label{sec:examples}
\begin{figure}[t]
    \centering
    \begin{tikzpicture}
    \newcommand\conf[2][]{
    \addplot[mark=none, color=#1]
        table [%
        col sep=comma,
        x=x, y expr={\thisrow{f}},] %
        {\currfiledir data/rbf/#2.csv};
    \addplot [name path=upper, mark=none, color=#1, dashed, forget plot]
        table [%
        col sep=comma,
        x=x, y expr={\thisrow{f} + \thisrow{std}},] %
        {\currfiledir data/rbf/#2.csv};
    \addplot [name path=lower, mark=none, color=#1, dashed, forget plot]
        table [%
        col sep=comma,
        x=x, y expr={\thisrow{f} - \thisrow{std}},] %
        {\currfiledir data/rbf/#2.csv};
    \addplot[color=#1, fill opacity=0.15, forget plot] fill between[of=lower and upper];
    }
        \begin{axis}[
        scale only axis, 
        axis lines=left,
        height=3cm, 
        width=8cm, 
        xmin=-1.1, 
        xmax=1.1, 
        ymin=-1.5, 
        ymax=1.5, 
        grid=both,
        legend entries={$p(f^*)$, $q_S(f^*)$, $q_I(f^*)$},
        legend columns=3,
        legend style={
        at={(0.5, 1.0)},
        anchor=south,
        draw=none
        }
        ]
        \conf[Paired-B]{base}
        \conf[Paired-D]{standard}
        \conf[Paired-H]{integral}
        \addplot +[mark=none, forget plot, name path=left] coordinates {(-0.5, -1.5) (-0.5, 1.5)};
        \addplot +[mark=none, forget plot, name path=right] coordinates {(0, -1.5) (0, 1.5)};
        \addplot[gray, fill opacity=.15] fill between[of=left and right];
        \node at (axis cs:-0.425,-1.05) {$\Omega$};
        \foreach \c in {-0.11, -0.33}
            {
                \addplot [color=Paired-H, samples=50, domain=\c-.175:\c+.175] {1.25 + 0.1*exp(-1/(0.1^2) * (x - \c)^2)};
            }
        \foreach \c in {-0.78, 0.78}
            {
                \addplot [color=Paired-D, samples=50, domain=\c-.175:\c+.175] {1.25 + 0.1*exp(-1/(0.1^2) * (x - \c)^2)};
            }
        \foreach \c in {-0.78, -0.55, -0.33, -0.11, 0.11, 0.33, 0.55, 0.78}
            {
            \addplot [color=Paired-B, samples=50, domain=\c-.175:\c+.175] { -1.4 + 0.1*exp(-1/(0.1^2) * (x - \c)^2)};
            }
        \end{axis}
    \end{tikzpicture}
    \caption{Posterior predictive distributions in an \gls{rbf} model. The full model is given by $p(f^*)$. The posteriors $q_I(f^*)$ and $q_S(f^*)$ use reduced models with bases selected by \cref{eq:Lbound,eq:simplifiedbound}, respectively. The \glspl{bf} for the original model are depicted at the bottom of the plot and the chosen \glspl{bf} for the reduced models are depicted at the top of the plot. Note that the proportions of the \glspl{bf} are exaggerated. The subdomain $\Omega$ of interest is highlighted in gray.}
    \label{fig:rbfselection}
\end{figure}
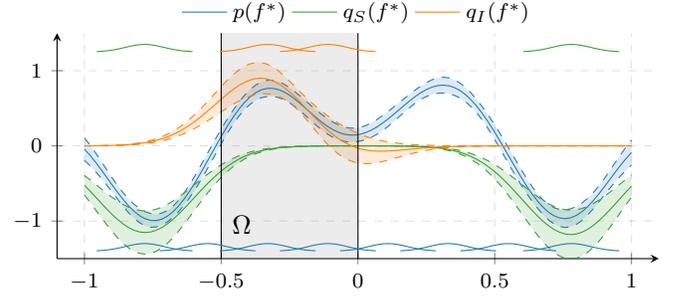

Our first numerical example considers selection in the case of \glspl{rbf}, as we, in this case, can intuitively determine which \glspl{bf} are relevant, confirming the validity of our approach.
Our second example considers a random function $f:\mathbb{R}^3\to\mathbb{R}$ drawn from a \gls{gp} prior. It is mainly used to illustrate the computational benefits of the \gls{bf} selection approach in models with thousands of parameters.
Lastly, a third example, where the method is applied to a real dataset for magnetic field mapping, is provided in the additional material.

\subsection{Radial basis functions}
We consider an \gls{rbf} expansion where the \glspl{rbf} are given by $\phi_i(x) = \exp\left( -\lVert x - c_i \rVert^2_2/l^2 \right)$, 
where $l$ and $c_i$ are parameters of \gls{bf} $i$.
We generate synthetic data from a function on the domain $[-1, 1]$ and place $L=10$ \glspl{bf} equidistantly spaced in the domain.
The posterior over the \gls{bf} weights $(m,S)$ is found through linear regression. 
We then select $|\mathcal{J}|=2$ \glspl{bf} for prediction in the subdomain $\Omega=[-0.5, 0]$ using both \cref{eq:Lbound,eq:simplifiedbound}.
The intuitive \gls{bf} choice is \glspl{bf} for which $c_i$ are close, in some sense, to $\Omega$, since the \glspl{bf} are exponentially decaying away from $c_i$; see \cite{visetSpatiallyScalableRecursive2023,kullbergOnlineJointState2021}.

The results are presented in \cref{fig:rbfselection}.
Clearly, the \gls{bf} selection using \cref{eq:Lbound} finds that the intuitively correct closest \glspl{bf} are the most important for prediction in $\Omega$.
The selection using \cref{eq:simplifiedbound}, on the other hand, identifies the \glspl{bf} on the edge of the domain as the most important.
This is natural, considering that, in this case, $\int_{\Omega} \lvert \phi_j(x)\rvert^2~dx \not\approx \int_{\Omega} \lvert \phi_i(x)\rvert^2~dx, \forall i, j$, violating the assumption of \cref{eq:simplifiedbound}.
Nevertheless, this confirms the validity of the \gls{bf} selection approach.

\subsection{Random function}
We consider a function $f:\mathbb{R}^3\to\mathbb{R}$ drawn from a \gls{gp} prior with zero mean and a squared-exponential kernel given by ${\kappa(x,x')=\sigma^2_f \exp \left(-\lVert x-x'\rVert^2/2l^2 \right)}$,
with lengthscale $l=0.1$ and variance $\sigma^2_f=0.05$.
The ${N=1000}$ training inputs are uniformly drawn from $[-1,1]^3$ and i.i.d. Gaussian noise with variance $\sigma^2=0.01$ is added to the outputs.
The baseline model is an \gls{hgp} defined on $[-2, 2]^3$. 
The \glspl{bf} are thus given by \cite{solinHilbertSpaceMethods2020}
\begin{equation*}
    \phi_i(\mathbf{x}) = \prod_{d=1}^3\frac{1}{\sqrt{2}}\sin\left(\frac{\pi i_d([\mathbf{x}]_d + 2)}{4}\right),
\end{equation*}
where $i$ is a multi-index and $i_d$ indexes each individual dimension.
The number of \glspl{bf} along each dimension is varied in $L_d\in[2, 20]$, corresponding to a total number of \glspl{bf} between $L=8$ and $L=8000$.

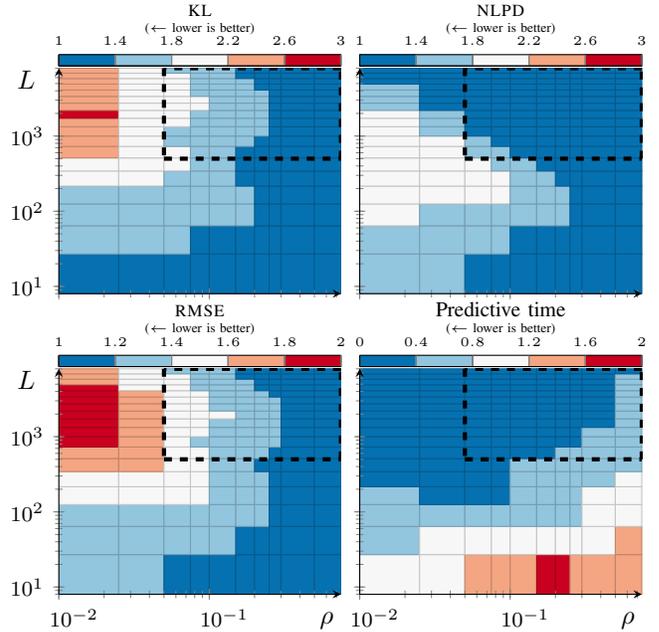
\begin{figure}[t]
    \centering
    \pgfplotsset{colormap/RdBu-5}
\begin{tikzpicture}%
\begin{groupplot}[
group style={group size=2 by 2, 
            horizontal sep=.25cm, 
            vertical sep=1cm,
            y descriptions at=edge left,
            x descriptions at=edge bottom,
},
height=3cm, 
width=3.75cm,
view={0}{90},
scale only axis,
colormap = {reverse RdBu-5}{
    indices of colormap={
        \pgfplotscolormaplastindexof{RdBu-5},...,0 of RdBu-5},
},
colormap access=const,
colorbar sampled,
colorbar horizontal,
colorbar style={
samples=5,
at={(0.5,1.01)},
anchor=south,
xticklabel pos=upper,
height=.15cm,
xtick=data,
xlabel={($\leftarrow$ lower is better)},
xlabel style={
    at={(0.5, 2.0)},
    font=\tiny,
},
ticklabel style={font=\tiny},
/pgf/number format/fixed,
/pgf/number format/precision=1,
},
title style={
anchor=south,
at={(0.5, 1.12)},
},
xmode=log, 
ymode=log, 
ylabel=$L$,
xlabel=$\rho$,
xlabel style={
    at={(0.95, -.04)},
},
ylabel style={
    at={(-.05, 0.95)},
    rotate=-90,
}
]

\nextgroupplot[
axis lines=left,
point meta min=1,
point meta max=3,
title=\textsc{kl},
]
    \addplot3[surf, 
        mesh/rows=11, 
        mesh/cols=19,
        shader=faceted,
        ] %
table[col sep=comma, x={x}, y={M}, z={kl},] {\currfiledir data/adaptiveselection.csv};
\addplot3[mesh,color=black, dashed, ultra thick] coordinates {(0.05, 500, 0) (0.75, 500, 0) (0.75, 8000 ,0) (0.05, 8000, 0) (0.05, 500, 0)};

\nextgroupplot[
axis lines=left,
title=\textsc{nlpd},
point meta min=1,
point meta max=3,
]
    \addplot3[surf, 
        mesh/rows=11, 
        mesh/cols=19,
        shader=faceted,
        ] %
table[col sep=comma, x={x}, y={M}, z={nlpd},] {\currfiledir data/adaptiveselection.csv};
\addplot3[mesh,color=black, dashed, ultra thick] coordinates {(0.05, 500, 0) (0.75, 500, 0) (0.75, 8000 ,0) (0.05, 8000, 0) (0.05, 500, 0)};

\nextgroupplot[
axis lines=left,
title=\textsc{rmse},
point meta min=1,
point meta max=2,
xticklabel style={xshift=.2cm},
]
    \addplot3[surf, 
        mesh/rows=11, 
        mesh/cols=19,
        shader=faceted,
        ] %
table[col sep=comma, x={x}, y={M}, z={rmse},] {\currfiledir data/adaptiveselection.csv};
\addplot3[mesh,color=black, dashed, ultra thick] coordinates {(0.05, 500, 0) (0.75, 500, 0) (0.75, 8000 ,0) (0.05, 8000, 0) (0.05, 500, 0)};

\nextgroupplot[
axis lines=left,
title=Predictive time,
point meta min=0,
point meta max=2,
xticklabel style={xshift=.2cm},
]
    \addplot3[surf, 
        mesh/rows=11, 
        mesh/cols=19,
        shader=faceted
        ] %
table[col sep=comma, x={x}, y={M}, z={time_min},] {\currfiledir data/adaptiveselection.csv};
\addplot3[mesh,color=black, dashed, ultra thick] coordinates {(0.05, 500, 0) (0.75, 500, 0) (0.75, 8000 ,0) (0.05, 8000, 0) (0.05, 500, 0)};

\end{groupplot}

\end{tikzpicture}
\caption{Performance of the reduced model relative to the full (original) model. $\rho$ is the fraction of \glspl{bf} chosen for prediction, \ie., $\rho=0.1$ chooses the $10\%$ most important \glspl{bf}. 
$L$ is the total number of basis functions available for the full model. 
The metrics are all relative, where blue is better and red is worse.}
\label{fig:hgp}
\end{figure}

To evaluate the quality of the approximate predictions, we consider the \gls{nlpd}, \gls{rmse}, and \gls{kl} divergence of the reduced model's and full model's predictive distributions on a dense grid of $N_{t}=3375$ test points.
The \gls{nlpd} and \gls{kl} are computed toward a standard \gls{gp} model with the same kernel.
Lastly, we benchmark the time necessary to compute the predictions.
For ease of presentation, the four metrics for the reduced and full model are then converted into ``relative'' metrics, by, \eg, $\frac{\gls{nlpd}_{\text{app.}}}{\gls{nlpd}_{\text{full}}}$.
All the metrics are plotted against the number of \glspl{bf} $L$ and fraction $\rho$ of retained \glspl{bf} in \cref{fig:hgp}.

The results are intuitively sound, as the approximation error grows as more \glspl{bf} are removed, \ie, for low values of $\rho$.
For low values of $L$, the method does not yield any speed-ups.
This is natural as the cost of the \gls{bf} selection process itself is then greater than the speed-up achieved by the reduction.
This is particularly apparent in the bottom right part of the relative predictive time plot, where the \gls{bf} selection process actually makes the predictions \emph{slower}.
However, already at $L\approx 200$, the \gls{bf} selection process starts speeding up the prediction process.
Considering all of the metrics, the sweet spot is around $L > 500$ and $\rho > 0.05$; see the black dashed area in \cref{fig:hgp}, where there is a low approximation error and low relative predictive time.
For $\rho\approx 0.3$, the \gls{kl}, \gls{nlpd}, and \gls{rmse} are all close to the full model, while the predictive time is greatly reduced as long as $L>200$.
Note that the optimal value of $\rho$ is problem-dependent.
However, as mentioned in \cref{sec:selection}, by selecting a proper threshold for the bounds \cref{eq:Lbound,eq:simplifiedbound}, they may be used to automatically select the number of \glspl{bf} to keep.
We leave this for future exploration.

\section{Conclusions}\label{sec:conclusions}

A method for a posteriori selecting relevant \gls{bf} in an already learned \gls{bf} expansion model has been proposed. 
The method enables model predictions to be calculated efficiently over subsets of the model domain with minor accuracy reduction. 
The method can be applied to a variety of \gls{bf} expansion models, such as \gls{rbf} expansions, sparse \gls{gp} approximations, etc., and enable these to tackle larger problems. 
An important application area for the proposed method is in multi-agent magnetic-field localization, where local and reduced-size \gls{gp}-based magnetic-field maps must be distributed to the agents.

\bibliographystyle{IEEEtran}
\bibliography{main}
\end{document}